\documentclass{naturemod}

\newif\ifnatureclass
\natureclasstrue
\usepackage{amsmath, amssymb, amsfonts, textcomp, graphicx, latexsym, psfrag, pstool, hyperref, bm, wasysym, xcolor,multibib}

\newcites{mthds}{Methods citations}

\usepackage{etoolbox}
\makeatletter
\patchcmd{\thebibliography}{%
  \section*{\refname}\@mkboth{\MakeUppercase\refname}{\MakeUppercase\refname}%
}{}{}{}
\makeatother

\ifnatureclass
\else
  
\fi
\usepackage{txfonts}

\hypersetup{
    colorlinks=true,       
    linkcolor=blue,          
    citecolor=blue,        
    filecolor=blue,      
    urlcolor=blue           
}

\def\Hz{\mbox{ Hz}}
\def\kHz{\mbox{ kHz}}
\def\nm{\mbox{ nm}}
\def\um{\mbox{ $\mu$m}}

\definecolor{darkred}{rgb}{0.5,0,0}
\definecolor{darkgreen}{rgb}{0,0.5,0}
\definecolor{darkblue}{rgb}{0,0,0.5}
\definecolor{darkcyan}{rgb}{0,0.5,0.5}
\definecolor{darkmagenta}{rgb}{0.5,0,0.5}
\definecolor{darkyellow}{rgb}{0.5,0.5,0}
\definecolor{darkpurple}{HTML}{562A7F}
\definecolor{darkbrown}{HTML}{753423}

\graphicspath{{./}{./figs/}}

\newcommand{\bq}{\begin{equation}}
\newcommand{\eq}{\end{equation}}
\newcommand{\bn}{\begin{eqnarray}}
\newcommand{\en}{\end{eqnarray}}

\begin{document}

\ifnatureclass
  
  \title{Hysteresis in a quantized superfluid atomtronic circuit}
  
  \author{Stephen Eckel$^{1}$, Jeffrey G. Lee$^{1}$, Fred Jendrzejewski$^{1}$, Noel Murray$^{2}$, Charles W. Clark$^{1}$, Christopher J. Lobb$^{1}$, William D. Phillips$^{1}$, Mark Edwards$^{2}$ \& Gretchen K. Campbell$^{1}$}
  
  \maketitle
  
  \begin{affiliations}
    \item Joint Quantum Institute, National Institute of Standards and Technology and University of Maryland, Gaithersburg, Maryland 20899, USA
    \item Department of Physics, Georgia Southern University, Statesboro, Georgia 30460-8031, USA
  \end{affiliations}

\else

  \title{Hysteresis in a quantized superfluid atomtronic circuit}

  \author{Stephen Eckel}
  \affiliation{Joint Quantum Institute, National Institute of Standards and Technology and University of Maryland, Gaithersburg, Maryland 20899, USA}
  \author{Jeffrey G. Lee}
  \affiliation{Joint Quantum Institute, National Institute of Standards and Technology and University of Maryland, Gaithersburg, Maryland 20899, USA}
  \author{Fred Jendrzejewski}
  \affiliation{Joint Quantum Institute, National Institute of Standards and Technology and University of Maryland, Gaithersburg, Maryland 20899, USA}
  \author{Noel Murray}
  \affiliation{Department of Physics, Georgia Southern University, Statesboro, Georgia 30460-8031, USA}
  \author{Charles W. Clark}
  \affiliation{Joint Quantum Institute, National Institute of Standards and Technology and University of Maryland, Gaithersburg, Maryland 20899, USA}
  \author{Christopher J. Lobb}
  \affiliation{Joint Quantum Institute, National Institute of Standards and Technology and University of Maryland, Gaithersburg, Maryland 20899, USA}
  \author{William D. Phillips}
  \affiliation{Joint Quantum Institute, National Institute of Standards and Technology and University of Maryland, Gaithersburg, Maryland 20899, USA}
  \author{Mark Edwards}
  \affiliation{Department of Physics, Georgia Southern University, Statesboro, Georgia 30460-8031, USA}
  \author{Gretchen K. Campbell}
  \affiliation{Joint Quantum Institute, National Institute of Standards and Technology and University of Maryland, Gaithersburg, Maryland 20899, USA}
  
  \maketitle
\fi

\ifnatureclass
\begin{abstract}
\else
{\bf
\fi
Atomtronics~\cite{Pepino2009,Beeler2013} is an emerging interdisciplinary field that seeks new functionality by creating devices and circuits where ultra-cold atoms, often superfluids,  play a role analogous to the electrons in electronics. 
Hysteresis is widely used in electronic circuits, e.g., it is routinely observed in superconducting circuits~\cite{Silver1967} and is essential in rf-superconducting quantum interference devices [SQUIDs]~\cite{Zimmerman1970}.
Furthermore, hysteresis is as fundamental to superfluidity~\cite{Mueller2002} (and superconductivity) as quantized persistent currents~\cite{Ramanathan2011,Moulder2012,Wright2013}, critical velocity~\cite{Onofrio2000,Inouye2001,Engels2007,Miller2007,Neely2010,Desbuquois2012},  and Josephson effects~\cite{Albiez2005,Levy2007}. Nevertheless, in spite of multiple theoretical predictions~\cite{Diakonov2002,Mueller2002,Watanabe2011,Morsch2006}, hysteresis has not been previously observed in any superfluid, atomic-gas Bose-Einstein condensate (BEC).
Here we demonstrate hysteresis in a quantized atomtronic circuit: a ring of superfluid BEC obstructed by a rotating weak link.
We directly detect hysteresis between quantized circulation states, in contrast to superfluid liquid helium experiments that observed hysteresis directly in systems where the quantization of flow could not be observed~\cite{Kojima1971} and indirectly in systems that showed quantized flow~\cite{Schwab1997,Schwab1998}.
Our techniques allow us to tune the size of the hysteresis loop and to consider the fundamental excitations that accompany hysteresis.
The results suggest that the relevant excitations involved in hysteresis are vortices and indicate that dissipation plays an important role in the dynamics. 
Controlled hysteresis in atomtronic circuits may prove to be a crucial feature for the development of practical devices, just as it has in electronic circuits like memory, digital noise filters (e.g., Schmitt triggers), and magnetometers (e.g., SQUIDs).
\ifnatureclass
\end{abstract}
\else
}
\fi

\begin{figure}
 \center
 \includegraphics{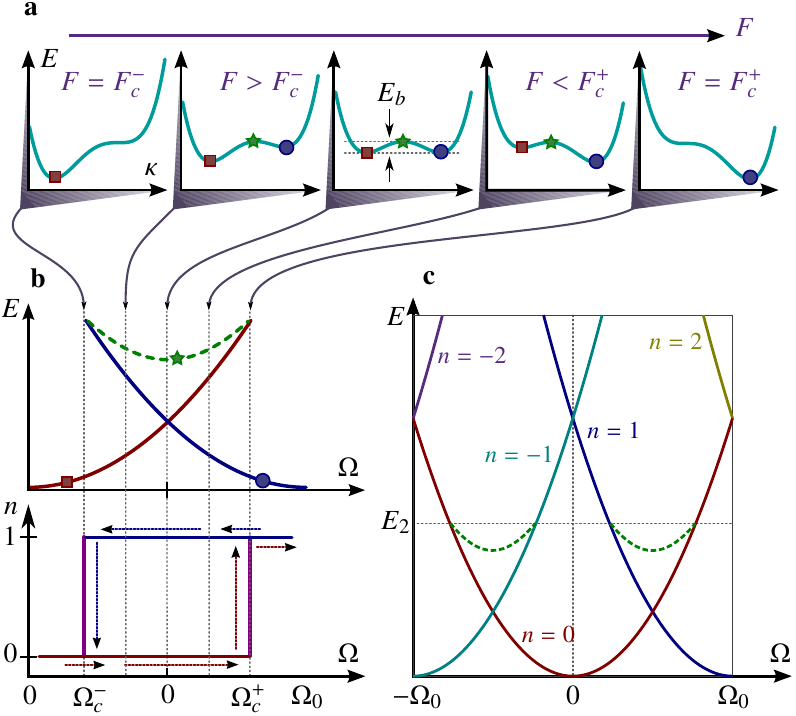}
 \caption{\label{fig:hyst_explanation} Origin of hysteresis. {\bf a}, A schematic of the energy landscape of a hysteretic system.  As a function of an order parameter $\kappa$, the energy can have local minima (\textcolor{darkred}{$\blacksquare$},\textcolor{darkblue}{\CIRCLE}), which represent stable states, separated by a local maximum (\textcolor{darkgreen}{$\bigstar$}), which forms an energy barrier $E_b$.  This landscape is shown for five values of the applied field $F$ (for superfluidity, $F\rightarrow\Omega$, the rotation rate of the trap). {\bf b}, Plotted as a function of $\Omega$ for a superfluid, the energy of the minima (solid) and maximum (dashed) form a swallowtail (upper), which exhibits hysteresis (lower). {\bf c}, This swallowtail structure is periodic in $\Omega_0$; states above $E_2$ are unstable.}
\end{figure}

Hysteresis is a general feature of systems where the energy has two (or more) local minima separated by an energy barrier.  A schematic of this type of energy landscape is shown in Fig~\ref{fig:hyst_explanation}a.  A canonical example of hysteresis is the Landau theory of ferromagnetism~\cite{PlischkeStatisticalPhysics}, where the order parameter $\kappa$ is the magnetization, and $E(\kappa)$ has two minima (stable states) corresponding to the magnetization being aligned or anti-aligned to the applied magnetic field.  In the case of a BEC in a ring-shaped trap, these minima represent stable flow states of the system, and their energies depend on the applied rotation rate of the trap, $\Omega$ (here, this rotation is created using a rotating repulsive perturbation). With no interatomic interactions, there is only one minimum in the energy landscape of the BEC.  With the addition of interactions, an energy barrier can appear creating two (or more) stable flow states.  This barrier stabilizes the flow, making the BEC a superfluid~\cite{Mueller2002,Baharian2013}.

The energy of the barrier is not generally known for superfluid systems; depending on the parameters of the system it could be related to the energy required to create elementary excitations such as phonons, solitons, or vortices.  On the other hand, the stable states are well known.   Rotation of a superfluid in a ring is characterized by a quantized rotation frequency $n\Omega_0$, where $n$ is the winding number, $\Omega_0=\hbar/(mR^2)$ is the rotational quantum, $\hbar$ is Planck's constant divided by $2\pi$, $m$ is the mass of an atom, and $R$ is the mean radius of the trap.  The energy of the superfluid in the frame that rotates with the trap depends on the relative velocity between the superfluid and the trap~\cite{Mueller2002,Baharian2013}, and the energy is proportional to $(n-\Omega/\Omega_0)^2$.

Any ring-shaped superfluid {\it must necessarily} exhibit both hysteresis and a critical rotation rate $\Omega_c^\pm$ (or, equivalently, a critical velocity), because all these effects fundamentally arise from the energy barrier that creates superfluidity.  To understand this, we plot the energy of the stable states and the energy barrier as a function of $\Omega$.   Fig.~\ref{fig:hyst_explanation}b shows this {\it swallowtail} energy structure.  If the system begins in $n=0$, the flow is stable until $\Omega=\Omega_c^+$, where the energies of the $n=0$ state and the barrier are equal.  At this point, the $n=0$ state is no longer stable and a transition occurs to $n=1$, which has lower energy.  If $\Omega$ is now decreased, this state is stable until $\Omega<\Omega_c^-$, where the flow changes.  Thus, a typical hysteresis loop is traced, as shown in the bottom of Fig.~\ref{fig:hyst_explanation}b.  Note that while the $\Omega_c^\pm$ are the same relative to the superfluid flow,  they are generally different in the lab frame and thus appear as hysteresis.  Furthermore, in the hysteretic case, the $\Omega_c^\pm$ are different from a more general definition of critical rotation (or velocity) that involves the onset of dissipation or the creation of excitations.  At $\Omega_c^\pm$, the hysteretic system may create excitations or experience dissipation, but both cease after the transition is made.   Measurement of a hysteresis loop, in addition to measuring $\Omega_c^\pm$, shows an important feature of the underlying energy landscape: the system has at least two stable states.  Bi-stablity of a moving BEC has been demonstrated independently of quantized states or critical velocities~\cite{Recati2001,Madison2001}.  Lastly, we note that unlike ferromagnetism, this energy structure is periodic in $\Omega$ with period $\Omega_0$, as shown in Fig.~\ref{fig:hyst_explanation}c.  Similar periodic swallowtails are predicted for superfluids trapped in a lattice~\cite{Morsch2006}.

\begin{figure}
 \center
 \includegraphics{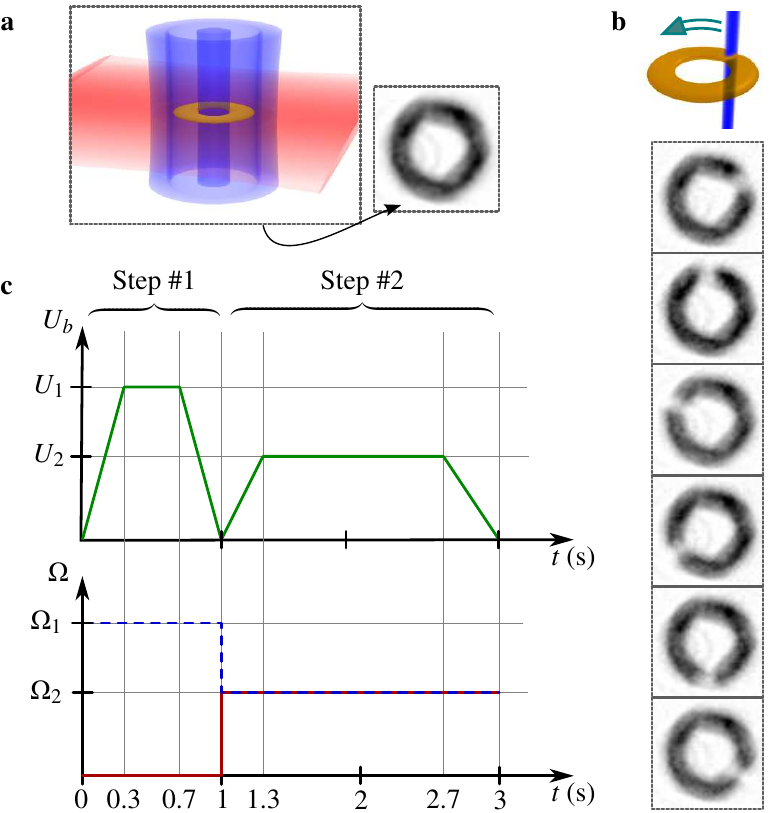}
 \caption{\label{fig:ring_images} Experimental setup and procedure. {\bf a}, Schematic and {\it in-situ} images of our trap, which is formed by crossing a ring-shaped dipole trap for radial confinement and a sheet trap for vertical confinement.  {\bf b}, Schematic and {\it in-situ} images of a ring stirred by a repulsive weak link.  {\bf c}, Two step experimental sequence: the height $U$ of the repulsive potential  and angular rotation rate $\Omega$ as a function of time.  Step \#1 sets the initial winding number using $\Omega_1$ (either $0\Hz$ or $1.1\Hz$) and $U_1$ ($\approx1.1\mu_0$); step \#2 probes the hysteresis with $\Omega_2$ and $U_2$ (see text).}
\end{figure}

Our superfluid system is a BEC of $^{23}$Na atoms in a ring-shaped optical dipole trap, as shown in Fig.~\ref{fig:ring_images}a.  To induce flow, a blue-detuned laser creates a rotating repulsive potential, depleting the density in a small portion of the ring and thereby creating a weak link~\cite{Hoskinson2006}.  The intensity of the laser sets the height of this potential, $U$.   Without this weak link, superfluid flow in the ring should be quite stable~\cite{Baharian2013} with $\Omega_c^+\gg\Omega_0$.   Changing $U$ will change the critical angular velocities $\Omega_c^\pm$ and the size of the hysteresis loop.  Rotating the weak link in the azimuthal direction at angular frequency $\Omega$, as shown in Fig.~\ref{fig:ring_images}b, can drive transitions, or phase slips, between the quantized circulation states~\cite{Wright2013}.

\begin{figure}
 \center
 \includegraphics{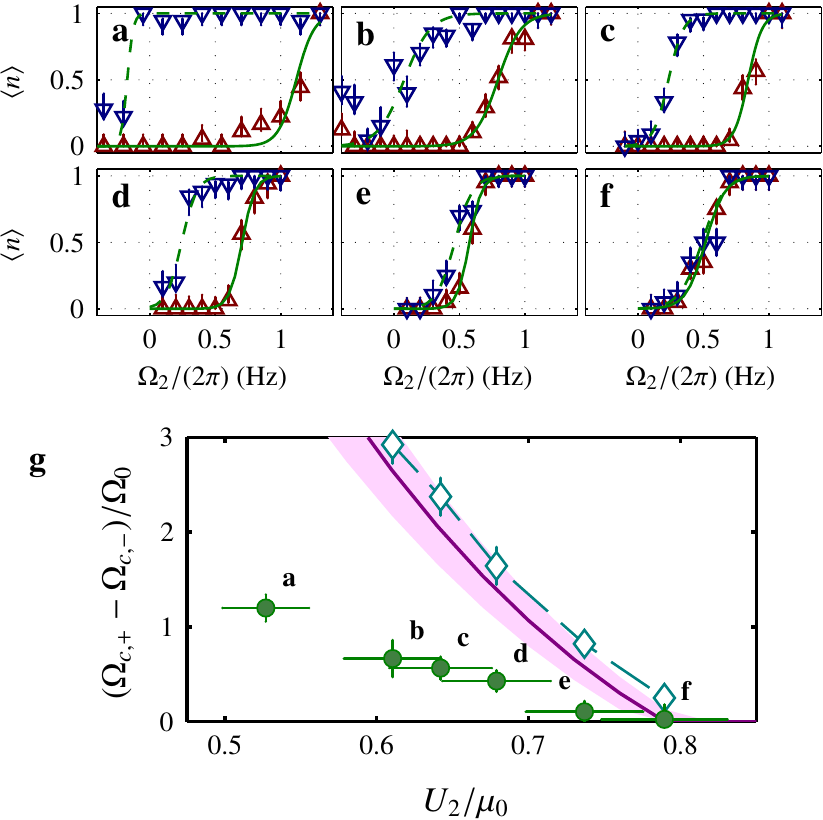}
 \caption{\label{fig:data} Hysteresis data.  {\bf a-f}, Hysteresis loops with sigmoid fits.  The red triangle (blue inverted triangles) show the winding number $n$ averaged over $\approx$20 shots, when starting with $n=0$ ($n=1$). All error bars show the 68\% confidence interval.  The fits determine $\Omega_c^\pm$ and $\Omega_0/2$ (gray, vertical lines; see Methods) and their uncertainties.  {\bf g}, Hysteresis loop size vs. $U_2$. The green circles show the experimental data.  The magenta line and band are the prediction and uncertainty of an effective 1D hydrodynamic model~\cite{Watanabe2009}.  The open (filled) cyan diamonds and their uncertainties are the results of our GPE simulation with $\Lambda=0$ ($\Lambda=0.01$).}
\end{figure}

To observe hysteresis in these phase slips, we use a two step experimental sequence, as shown in Fig.~\ref{fig:ring_images}c.  After condensing the atoms into the ring trap, the BEC is prepared into either $n=0$ or $n=1$ circulation states by either not rotating the weak link or by rotating it at $\Omega_1=1.1$~Hz.  The fidelity with which this procedure generates the expected initial state is $\gtrsim97\%$.  We then stir the weak link at various angular velocities $\Omega_2$ for an additional 2~s.  $\Omega_2$ spans the range between $-0.3\Hz$ and $1.2\Hz$.  In step \#1, $U$ is ramped to $U_1\approx1.1\mu_0$, where $\mu_0$ is the global chemical potential.  In step \#2, $U$ is ramped to a chosen $U_2$.  The transitions from $n=0\rightarrow1$ and $n=1\rightarrow0$ occur at different values of $\Omega_2$ and form hysteresis loops, as  Fig.~\ref{fig:data}a--f show.   Each plot shows the measured hysteresis loop for a specific $U_2$.  As $U_2$ is increased, both $\Omega_c^+$ and $\Omega_c^-$ become closer to $\Omega_0/2$, i.e., the hysteresis loop becomes smaller.  The observed transitions are not sharp unlike those in Fig.~\ref{fig:hyst_explanation}b.  The dominant broadening mechanism is likely shot-to-shot atom number fluctuations, but the non-zero temperature ($\approx100$~nK) may also contribute (see Supplemental Material). 

Fig.~\ref{fig:data}g shows the measured size of the hysteresis loop, $(\Omega_c^+-\Omega_c^-)/\Omega_0$, as a function of the strength of the weak link; the size of the loop monotonically decreases with increasing $U_2/\mu_0$ until it reaches a value consistent with zero near $U_2/\mu_0\approx0.75$.  To predict the size of the hysteresis loop, we used two models.  First, we used an effective one-dimensional model which, as a function of $\Omega_2$, computes the fluid velocity in the rotating frame.  We assume that $\Omega_c^\pm$ will occur when this velocity reaches the local speed of sound~\cite{Watanabe2009}.  As a separate approach, we also simulated our system with the 3-D, time-dependent Gross-Pitaevskii equation (GPE).  These two approaches predict hysteresis and are consistent, suggesting that both theories predict that $\Omega_c^\pm$ is determined by the sound speed.   Despite occurring at the sound speed, the observed excitations in the GPE simulation are vortex/anti-vortex pairs.  Perhaps most strikingly, Fig.~\ref{fig:data}g shows a large discrepancy between our models and experiment.

One property of the system that our models fail to include is dissipation.  As another approach, we added dissipation to the GPE phenomenologically~\cite{Choi1998},
\begin{equation}
 i\hbar\frac{\partial\psi}{\partial t} = (1-i\Lambda)\left[-\frac{\hbar^2}{2m}\nabla^2+ V(x,y,z,t) + g N |\psi|^2 - \mu\right]\psi\ ,
\end{equation}
where $\psi$ is the BEC wavefunction, $g$ is the interaction strength, $V$ is the externally applied potential (trap and weak link), $N$ is the atom number, $\mu$ is the chemical potential of the initial stationary state, and $\Lambda$ is the dissipation parameter.   With $\Lambda=0.01$, a reasonable value for our experiment, the hysteresis loop size decreases as shown in Fig.~\ref{fig:data}g but not  significantly compared to the discrepancy with experiment.   Increasing the damping parameter further does not improve the agreement (see Supplemental Material).  However, it is clear that dissipation is important.  In fact, dissipation is essential and implicitly assumed in the energy landscape picture described in Fig.~\ref{fig:hyst_explanation}: dissipation allows the system to relax to the minima of the landscape; without dissipation, the system cannot change its energy.

\begin{figure}
 \center
 \includegraphics{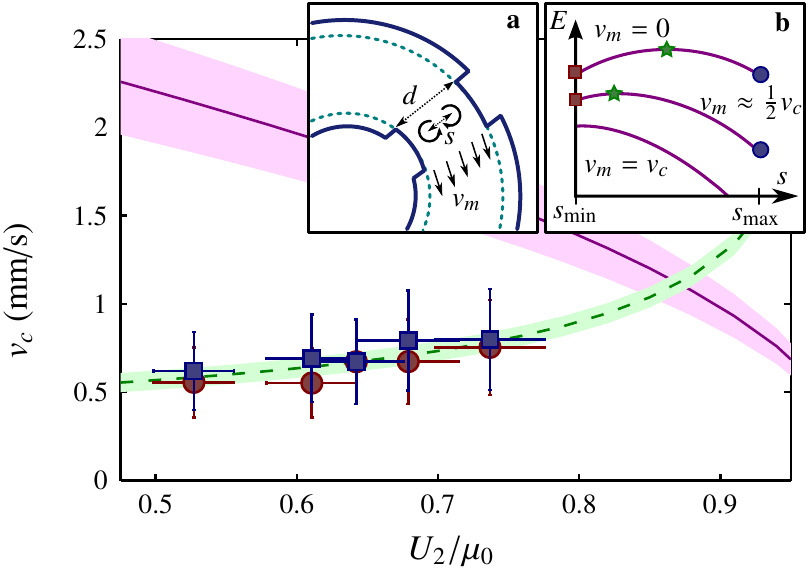}
 \caption{\label{fig:critical_velocity}  Extracted critical velocities vs. $U_2$.  The red circles (blue squares) show the critical velocities extracted from $\Omega_c^+$ ($\Omega_c^-$).  The magenta line and band show the estimate and uncertainty of the local speed of sound.  The green, dashed line and band show the best fit of the toy model of vortex creation and its statistical uncertainty.  All uncertainties are $1\sigma$.  {\bf a},  A diagram of a vortex/anti-vortex pair in a weak link of width $d$ and a vortex/anti-vortex separation $s$.  {\bf b}, Energy landscape as a function of $s$ for three different values of the velocity in the weak link region $v_m$, showing the stable states (\textcolor{darkred}{$\blacksquare$},\textcolor{darkblue}{\CIRCLE}) and the energy barrier (\textcolor{darkgreen}{$\bigstar$}).}
\end{figure}

To gain insight, we consider a toy model where the relevant excitations are vortex/anti-vortex pairs and derive the associated energy landscape.  If a (anti-)vortex were to be nucleated at the (inner) outer edges (as shown in Fig.~\ref{fig:critical_velocity}a), move to center, and annihilate, the winding number would change by one unit.  The energy of a vortex/anti-vortex pair in a perfectly hard-walled ring trap in the presence of a velocity field has been derived using the method of images~\cite{Fetter1967}.  In the limit that the width of the annulus $d\ll R$, this energy reduces to
\bq
 \label{eq:vortex_energy}
 E = \pi \rho d R \varv^2  + 2\pi\frac{\hbar \rho s \varv}{m} + 2\pi\frac{\rho \hbar^2}{m^2}\ln\left[\frac{d}{\pi\xi}\sin\left(\frac{\pi s}{d}\right)\right]\ ,
\eq
where $s$ is the separation between the vortices, $\varv$ is the velocity of the superfluid, $\rho$ is the effective 2D mass density, and $\xi$ is the healing length of the condensate and therefore the core size.  This equation applies to a system with a uniform annulus width $d$ and uniform velocity $\varv$.  To apply this model to our system, we take $d$ to be the effective width of the annulus in the weak link region and $\varv=\varv_m$, the maximum velocity in the weak link.  For $\varv_m=0$, Eq.~\ref{eq:vortex_energy} has a maximum at $s=d/2$ while diverging negatively at $d$ and $0$.  Such a divergence is unphysical, because the non-zero radii of the vortices prevent them from coming arbitrarily close to each other or the wall.  We assume the distance of closest approach to the walls is $C\xi$ and between vortices is $2C\xi$, where $C$ is of order unity.  Thus, $s$ ranges from $s_\text{min} = 2 C\xi$ to $s_\text{max} = d - 2 C \xi$.  We assume that vortices annihilate at $s_\text{min}$ and enter the annulus at $s_\text{max}$.

We plot the energy landscape described above in Fig.~\ref{fig:critical_velocity}b for several different $\varv_m$ and constant $d$.  The two stable states, at $s_\text{min}$ and $s_\text{max}$, represent a winding number difference of one.  This implies that for a phase slip to occur the vortex pair must nucleate at either $s_\text{max}$ or $s_\text{min}$ and move to the opposite extreme.  This happens when the energy barrier disappears, {\it i.e.}, when $dE/ds|_{s=s_\text{min(max)}} = 0$.  This defines the critical velocity $\varv_m=\varv_c$ as 
\bq
 \label{eq:critical_velocity}
 \varv_c = \pm\frac{\pi \hbar}{m d}\cot\left(\pi\frac{2 C \xi}{d}\right)\ ,
\eq
where $+$($-$) refers to starting at $s_\text{min}$($s_\text{max}$).

To compare this model to our experiment, we computed the critical velocity in the weak link from the transitions in the hysteresis loops.  The critical velocity is not a simple function of $U_2/\mu_0$ and $\Omega_2$; rather the requirements of quantized winding number and, in a frame co-rotating with the weak link, continuity of flow, require a self-consistent solution for the flow velocity around the entire ring (see Methods).  Fig.~\ref{fig:critical_velocity} shows the result of this calculation.  In direct contrast to the local speed of sound in the weak link, which decreases as $\sqrt{1-U_b/\mu_0}$, we find that the critical velocity {\it increases}.   The observed critical velocity is well fit to Eq.~\ref{eq:critical_velocity} with a single value of $2 C\xi/d$.  This value implies a distance of closest approach $2 C\xi \approx 0.4 d$.  Over the region of interest, the value of $C$ ranges from 1.5 to 0.7, agreeing with the assumption that it is of order unity ($C$ is calculated using the best estimates of $\xi$ and $d$, both of which vary with $U$).  The fact that the data can be fit using this crude model suggests that vortices are the relevant excitations and Eq.~\ref{eq:vortex_energy} (or something that captures similar physics) gives a good prediction of the energy landscape.

Our hysteretic system has the essential features of the rf-SQUID.  Just as SQUIDs detect magnetic fields, our analogous system can detect rotations.  Hysteresis plays an important role in rf-SQUIDs, where it is used as a readout mechanism.  In our system, hysteresis will also be important, allowing for greater accuracy by canceling systematic effects.  The hysteresis loops are centered about $\pm\Omega_0/2$ for different directions of rotation; therefore, one can measure the asymmetry in the measurements of $\pm\Omega_0/2$ to extract an unknown bias rotation.  Such measurements may cancel out effects like asymmetries in the ring potential.

In conclusion, we have measured hysteresis in a dilute atomic gas BEC, a phenomenon that is as fundamental to superfluidity as the existence of persistent currents and critical velocities.  Our studies suggest that the elementary excitations involved in hysteresis are vortices and that dissipation plays an important role in the dynamics.   We suspect that more sophisticated models that include dissipation will yield better agreement.  Finally, beyond being an atomtronic rotation sensor, it is possible that in the hysteretic regime this device could act as classical memory or a digital noise filter in future atomtronic circuits.

\subsection{Methods Summary}
\ifnatureclass
{
\else
{\small
\fi
The ring-shaped BEC, which contains approximately $4\times10^5$ $^{23}$Na atoms, is created from a cloud of laser cooled atoms by evaporation, first in a magnetic trap and then in a ring-shaped optical dipole trap.  The optical dipole trap is shaped roughly like a washer (see Methods), with measured harmonic trap frequencies in the vertical direction of $472(4)\Hz$ and 188(3)~Hz in the radial direction.  (Uncertainties in this paper are the uncorrelated combination of $1\sigma$ statistical and systematic uncertainties unless stated otherwise.)  The mean radius of the trap is $19.5(4)\um$.  The weak link is created by a blue-detuned laser beam (see Methods).  Time of flight expansion of the condensate allows us to determine the winding number by measuring the size of the central hole size that appears in the cloud.
}

\bibliographystyle{naturemagmod}
\bibliography{NaBEC_library}

\ifnatureclass
\begin{addendum}
\item [Supplementary Information] is available in the online version of the paper.
\item[Acknowledgements] This work was partially supported by ONR, the ARO atomtronics MURI, NIST, the  NSF through the PFC at the JQI and grant PHY-1068761. S.E. is supported by a National Research Council postdoctoral fellowship. We wish to thank K. Wright, W. T. Hill, III, and A. Kumar for valuable discussions and experimental assistance.
\item[Author Contributions] S.E, J.G.L and F.J took the experimental data. N.M, C.W.C and M.E developed and performed the GPE simulations. All authors were involved in analysis and discussions of the results, and contributed to writing the manuscript.
\item[Author Information]Reprints and permissions information is available at www.nature.com/reprints. The authors declare no competing financial interests. Correspondence and requests for materials should be addressed to G.K.C. (gretchen.campbell@nist.gov).
\end{addendum}
\fi

\ifnatureclass
\begin{methods}
\else
\section{Methods}
{\small
\fi
\ifnatureclass  
\subsection{Optical dipole traps.} 
\else
{\bf Optical dipole traps.} 
\fi
Our optical dipole trap is formed by the combination of two laser beams.  A blue-detuned ($\lambda = 532\nm$) laser beam passes through a ring-shaped intensity mask, and the shadow is imaged onto the atoms forming a repulsive, ring-shaped potential.  This trap combines with an attractive confining potential in the vertical direction, generated by a red-detuned ($\lambda = 1064\nm$) laser beam shaped like a sheet.  If the imaging resolution were perfect, the trap would be hard-walled in the radial direction, but it is in fact closer to a Gaussian with $1/e^2$ radius of $8.9(9)\um$.

To create the weak link, an acoustic optical deflector elongates a blue-detuned, focused, Gaussian beam by scanning radially at 2~kHz.  The beam was turned on and off with a 300~ms linear ramp.  The $1/e^2$ half-width of the weak link along the azimuthal direction is approximately $6\um$, limited by the resolution of our imaging system.   The size of the weak link along the radial direction is $\approx 50\%$ larger than the Thomas-Fermi width of the BEC.

We calibrated the weak link by observing the atomic density depletion caused by the weak link potential (see Supplemental Material).  The dominant uncertainty in $U_2/\mu_0$ is in the common calibration and is reflected in the horizontal error bars in Figs.~\ref{fig:data} \& \ref{fig:critical_velocity}; the relative uncertainties between the points are smaller.

\ifnatureclass  
\subsection{BEC parameters.}
\else
{\bf BEC parameters.}
\fi
 Approximately $4\times10^5$ atoms comprise the BEC after evaporation first in a magnetic time orbiting potential (TOP) trap and subsequently in the optical traps described above.  We estimate the global chemical potential $\mu_0$ to be $\mu_0/\hbar\approx2\pi\times(1.7\kHz)$ and the corresponding Thomas-Fermi full width in the vertical (radial) direction to be $3.2\um$ ($8.1\um$). Given this mean radius, we expect $\Omega_0 = 1.19(4)\Hz$, in rough agreement with the measured value of $\Omega_0=1.05(5)\Hz$ (by assuming the hysteresis loops are centered on $\Omega_0/2$).

\ifnatureclass  
\subsection{Measurement of the winding number.}
\else
{\bf Measurement of the winding number.}
\fi
 To measure the final rotational state after stirring, the BEC is released from the trap and imaged after 10~ms of time-of-flight (TOF).  As the BEC expands, rotation will cause a hole to appear in the center.  As with the winding number, the size of this feature is quantized and enables determination of the final circulation state~\cite{Wright2013,Moulder2012}.  Direct release from the repulsive dipole trap does not allow the hole to be resolved in 10~ms TOF, so we transfer first to an attractive ring and apply a decompression procedure similar to that of Wright, {\it et al.}~\cite{Wright2013}, before using partial transfer absorption imaging~\citemthds{Ramanathan2012}.

\ifnatureclass  
\subsection{Estimating the uncertainty in the average winding number.} 
\else
{\bf Estimating the uncertainty in the average winding number.} 
\fi
 Given that the outcome of any given experiment is either $n=0$ or $n=1$, traditional methods  of estimating the uncertainty in the mean value (e.g., Gaussian statistics) are not applicable.  The uncertainty in this average can be estimated by the cumulative beta distribution, which is appropriate for experiments that yield binary results~\citemthds{Cameron2011}.

\ifnatureclass  
\subsection{Fitting the hysteretic transitions and determining $\Omega_c^\pm$ and its uncertainty.}
\else
{\bf Fitting the hysteretic transitions and determining $\Omega_c^\pm$ and its uncertainty.}
\fi
 We use a sigmoid of the form $1/[\exp\{-(\Omega_2-\Omega_t)/\delta\Omega\}+1]$ to fit the data as in Fig.~\ref{fig:data}, where $\Omega_t$ and $\delta\Omega$ are the fit parameters.  While this fit well describes the data, the relationship between $\Omega_c^\pm$ and the fit parameters depends on the mechanisms for the broadening of the transition.  For example, consider a model where thermal fluctuations drive the system over the energy barrier.  This would occur when the energy barrier becomes of the order of $k_B T$, where $k_B$ is the Boltzmann constant and $T$ is the absolute temperature.  The dynamics of this process are random and would lead to phase slips at lower values of $\Omega_2$ for $0\rightarrow1$ transition (and higher values for $1\rightarrow0$ transitions).  In principle, this effect would cause a broadening of the transition region, and the zero temperature $\Omega_c^\pm$ would then correspond to the value of $\Omega_2$ where probability for a transition equals unity.  However, a different mechanism could be responsible for the broadening. In particular, atom number fluctuations can change $U_2/\mu_0$, and therefore $\Omega_c^\pm$, from shot to shot.  On average, this leads to a broadening.  Based on the experimentally observed change in the $\Omega_c^\pm$ vs. the strength of the weak link and our atom number shot to shot fluctuations of $\approx16$\% (this represents the peak-to-peak fluctuations for 95\% of the data), we expect the transitions to be approximately $0.12\Hz$ wide, compared to the average of 0.18~Hz.  Because atom number fluctuations explain most of the width, we take $\Omega_c^\pm = \Omega_t$, and, to account for the possibility of finite temperature or other unknown broadening effects, take the 1$\sigma$ uncertainty to be $\frac{3}{2}\delta\Omega$.

\ifnatureclass  
\subsection{Extracting the critical velocity.} 
\else
{\bf Extracting the critical velocity.} 
\fi
Extracting the critical velocity in our system is non-trivial because the flow must satisfy the requirements of quantized winding number and continuity of flow in the frame rotating with the weak link.  (Continuity of flow does not occur in any other frame.)  One counterintuitive result of these requirements is that moving the weak link will impart some angular momentum to the superfluid as viewed from the fixed, laboratory frame even in the $n=0$ state~\citemthds{Kojima1971,LambHydrodynamics}.
To extract the critical velocity given these constraints, we work in the rotating frame.  The velocity $v_r$ of atoms in the rotating frame is related to the rotation rate of the weak link by 
\bq
 \label{eq:vel_det}
 \frac{m}{\hbar}\int_0^{2\pi} v_r(\theta)\ Rd\theta + 2\pi\frac{\Omega}{\Omega_0} = 2\pi n\ ,
\eq
where $\theta$ is the azimuthal angle.  This equation is an expression of the Bohr-Sommerfeld quantization condition.  The first term represents the phase accumulated by the atoms after integrating once around the ring and the second term represents the Sagnac (Peierls) phase that appears due to transforming into the rotating frame.  In the rotating frame, the velocity $v_r(\theta)$ and the mass density $\rho(\theta)$ satisfies a continuity equation: $\rho(\theta_1)v_r(\theta_1) = \rho(\theta_2)v_r(\theta_2)$, where $\theta_1$ and $\theta_2$ are any two azimuthal angles.  Given a $U_2/\mu_0$, we determine the equivalent 1D density $\rho(\theta)$ by integrating over the radial and vertical directions of our cloud using the Thomas-Fermi approximation.  For a given rotation rate $\Omega$ and density $\rho(\theta)$, Eq.~\ref{eq:vel_det} and the continuity equation determine $v_r(\theta)$ and, in particular, the velocity in the weak link $v_m = \text{max}[v_r(\theta)]$ uniquely. The critical velocity is then taken to be the value of $v_m$ when the weak link is rotated at the critical rotation rate $\Omega_c^\pm$.

When $\Omega_c^+ - \Omega_c^-\rightarrow0$, this method of extracting $v_c$ is unreliable and thus we neglect the point near $U\approx0.8\mu_0$ (see Supplemental information).  Going slightly further into the regime where $U>\mu_0$ results in the BEC being broken.
\ifnatureclass
\end{methods}
\else
}
\fi

\bibliographystylemthds{naturemagmod}
\bibliographymthds{NaBEC_library}

\end{document}